\documentclass[twocolumn]{aastex61}

\usepackage{color}
\usepackage{graphicx}
\usepackage{subfigure}
\usepackage{topcapt}
\usepackage{float}
\usepackage{natbib}
\usepackage{amsmath}

\graphicspath{ {figs/} }

\newcommand{\nustar} {\textit{NuSTAR}}
\newcommand{\rhessi} {\textit{RHESSI}}

\newcommand{\angstrom}{\mbox{\normalfont\AA}}

\begin{document}

\title{First {\nustar} Limits on Quiet Sun Hard X-Ray Transient Events}

\correspondingauthor{Andrew J. Marsh}
\email{anjmarsh@ucsc.edu}
\author{Andrew J. Marsh}
\affiliation{Santa Cruz Institute for Particle Physics and Department of Physics, University of California, Santa Cruz, CA 95064, USA}
\author{David M. Smith}
\affiliation{Santa Cruz Institute for Particle Physics and Department of Physics, University of California, Santa Cruz, CA 95064, USA}
\author{Lindsay Glesener}
\affiliation{School of Physics \& Astronomy, University of Minnesota Twin Cities, Minneapolis, MN 55455, USA}
\author{Iain G. Hannah}
\affiliation{SUPA School of Physics \& Astronomy, University of Glasgow, Glasgow G12 8QQ, UK}
\author{Brian W. Grefenstette}
\affiliation{Cahill Center for Astrophysics, 1216 E. California Blvd, California Institute of Technology, Pasadena, CA 91125, USA}
\author{Amir Caspi}
\affiliation{Southwest Research Institute, Boulder, CO 80302, USA} 
\author{S\"{a}m Krucker}
\affiliation{Space Sciences Laboratory University of California, Berkeley, CA 94720, USA}
\affiliation{University of Applied Sciences and Arts Northwestern Switzerland, 5210 Windisch, Switzerland}
\author{Hugh S. Hudson}
\affiliation{SUPA School of Physics \& Astronomy, University of Glasgow, Glasgow G12 8QQ, UK}
\affiliation{Space Sciences Laboratory University of California, Berkeley, CA 94720, USA}
\author{Kristin K. Madsen}
\affiliation{Cahill Center for Astrophysics, 1216 E. California Blvd, California Institute of Technology, Pasadena, CA 91125, USA}
\author{Stephen M. White}
\affiliation{Air Force Research Laboratory, Space Vehicles Directorate, 3550 Aberdeen Ave SE, Kirtland AFB, NM 87117, USA}
\author{Matej Kuhar}
\affiliation{University of Applied Sciences and Arts Northwestern Switzerland, 5210 Windisch, Switzerland}
\affiliation{Institute for Particle Physics, ETH Z\"urich, 8093 Z\"urich,
Switzerland}
\author{Paul J. Wright}
\affiliation{SUPA School of Physics \& Astronomy, University of Glasgow, Glasgow G12 8QQ, UK}
\author{Steven E. Boggs}
\affiliation{Space Sciences Laboratory University of California, Berkeley, CA 94720, USA}
\author{Finn E. Christensen}
\affiliation{DTU Space, National Space Institute, Technical University
of Denmark, Elektrovej 327, DK-2800 Lyngby, Denmark}
\author{William W. Craig}
\affiliation{Space Sciences Laboratory University of California, Berkeley, CA 94720, USA}
\affiliation{Lawrence Livermore National Laboratory, Livermore, CA
94550, USA}
\author{Charles J. Hailey}
\affiliation{Columbia Astrophysics Laboratory, Columbia University, New York, NY 10027, USA}
\author{Fiona A. Harrison}
\affiliation{Cahill Center for Astrophysics, 1216 E. California Blvd, California Institute of Technology, Pasadena, CA 91125, USA}
\author{Daniel Stern}
\affiliation{Jet Propulsion Laboratory, California Institute of Technology, 4800 Oak Grove Drive, Pasadena, CA 91109, USA}
\author{William W. Zhang}
\affiliation{NASA Goddard Space Flight Center, Greenbelt, MD 20771, USA}

\begin{abstract}
We present the first results of a search for transient hard X-ray (HXR) emission in the quiet solar corona with the  \textit{Nuclear Spectroscopic Telescope Array} ({\nustar}) satellite. While {\nustar} was designed as an astrophysics mission, it can observe the Sun above 2~keV with unprecedented sensitivity due to its pioneering use of focusing optics. {\nustar} first observed quiet Sun regions on 2014 November 1, although out-of-view active regions contributed a notable amount of background in the form of single-bounce (unfocused) X-rays. We conducted a search for quiet Sun transient brightenings on time scales of 100 s and set upper limits on emission in two energy bands. We set 2.5--4~keV limits on brightenings with time scales of 100 s, expressed as the temperature T and emission measure EM of a thermal plasma. We also set 10--20~keV limits on brightenings with time scales of 30, 60, and 100 s, expressed as model-independent photon fluxes. The limits in both bands are well below previous HXR microflare detections, though not low enough to detect events of equivalent T and EM as quiet Sun brightenings seen in soft X-ray observations. We expect future observations during solar minimum to increase the {\nustar} sensitivity by over two orders of magnitude due to higher instrument livetime and reduced solar background.
\end{abstract}


\section{Introduction}
Hard X-rays (HXRs) are an important probe of particle acceleration and heating in solar flares. High-temperature plasma emission ($>$1 MK) can be seen directly via thermal processes (bremsstrahlung, free-bound continua, and emission lines). Many flare observations also show non-thermal bremsstrahlung distributions above $\sim$10 keV; these spectra can be inverted to give information about the underlying electron spectra. In many flares the energy in accelerated electrons and ions is comparable to the total radiated energy at all wavelengths \citep{Lin1976, Ram1995, Ems2012}. Therefore in order to fully understand the physical processes underlying solar flares, HXR measurements are necessary. 

The \textit{Reuven Ramaty High Energy Solar Spectroscopic Imager} ({\rhessi}) is the only currently dedicated solar HXR satellite \citep{Lin2002}. {\rhessi} can observe flares ranging in size from \textit{GOES} A-class microflares to the largest X-class events, due to movable shutters that reduce the measured flux below a certain photon energy threshold. \citet{Han2008} and \citet{Chr2008} showed that even the smallest detectable {\rhessi} events have characteristics similar to larger flares: they occur in active regions, show thermal emission from loops, and show impulsive, non-thermal emission from footpoints. It is an open question whether HXR-emitting flares exist outside of active regions, as {\rhessi} is unable to measure flux from the quiet Sun due to limited sensitivity and dynamic range \citep{Han2010}. 

Flares, or flare-like brightenings, contribute to the heating of the solar corona. \citet{Hud1991} showed that for a distribution of flare frequency versus energy, the smallest events dominate energetically if the power-law index is $>$2. Observations show a power-law index below this limit, but an exact value is difficult to determine due to selection bias and the use of different instruments at different energies. While ordinary flares do not provide enough energy to heat the corona \citep{Shi1995, Han2011}, it is possible that many undetected weaker events might.

\begin{figure*}[ht]
  \centering
  \begin{tabular}{ccc}
    \includegraphics[width=\columnwidth]{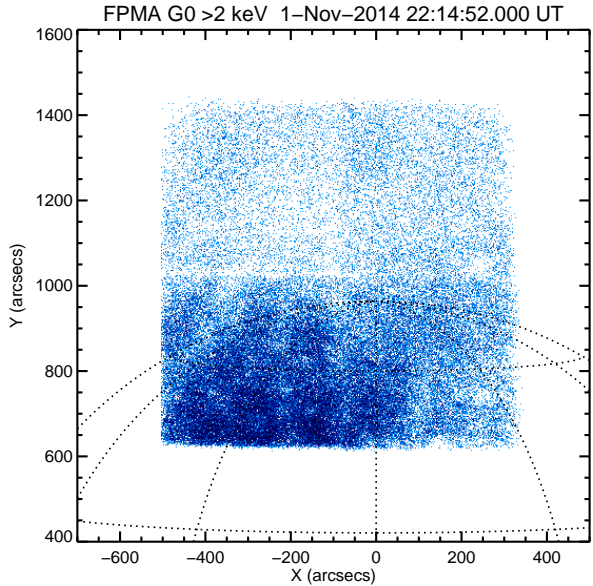}
    \includegraphics[width=\columnwidth]{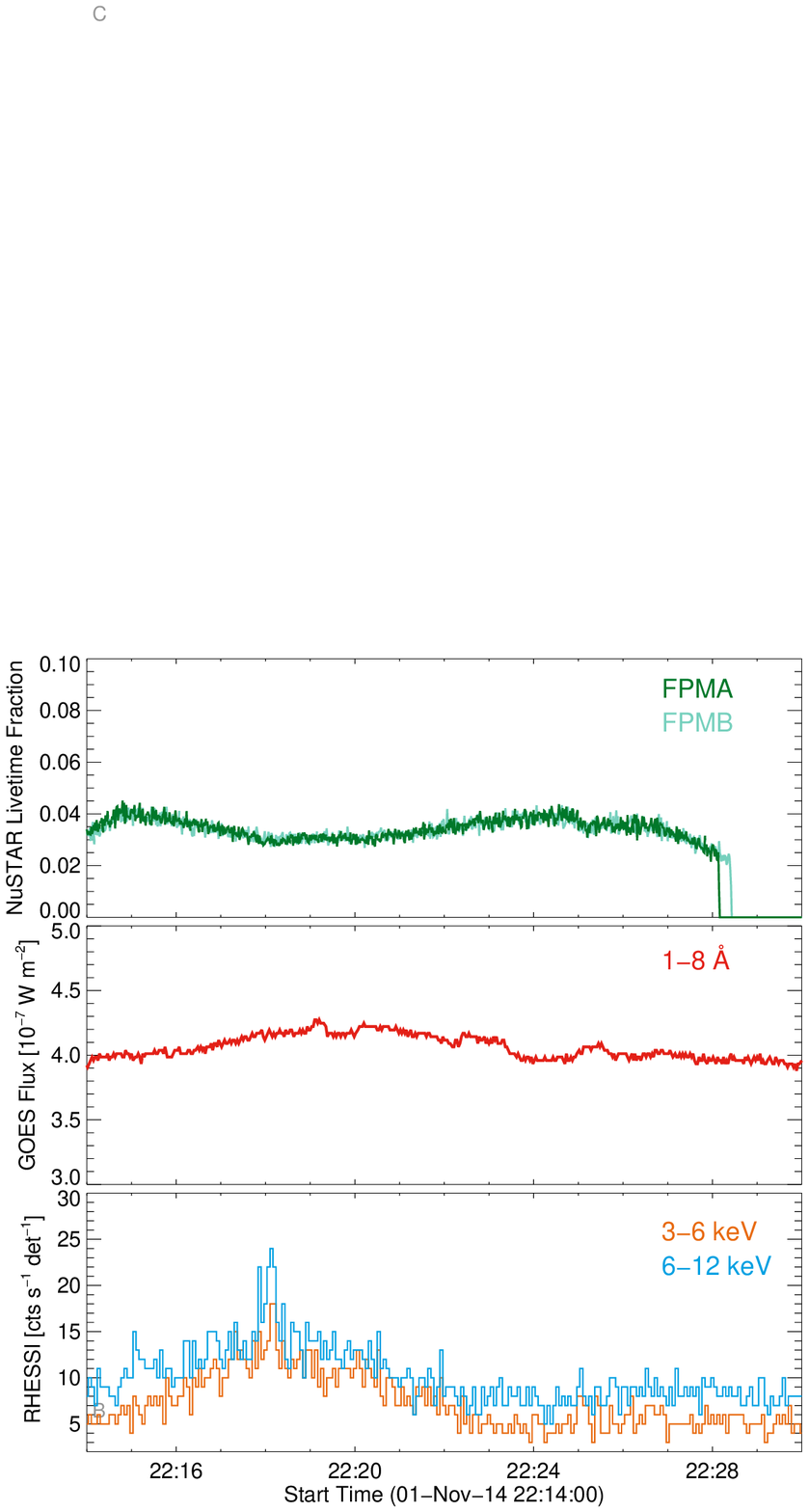}
  \end{tabular}
  \caption{(Left) {\nustar} image $>$2~keV in the FPMA telescope integrated over the 2014 November 1 north pole pointing. The detected emission is consistent with ghost rays produced by active regions outside the instrument FoV. (Right) Time profiles of the {\nustar} livetime (top panel), the \textit{GOES} 1--8~$\angstrom$ flux (middle panel), and the {\rhessi} 3--6 and 6--12~keV fluxes (bottom panel). The slow rise peaking at 22:18 UT in the {\rhessi} lightcurve is solar in origin, but outside {\nustar}'s FoV.}
  \label{fig:profiles}
\end{figure*}

Quiet-Sun transient brightenings (also referred to in the literature as heating events, network flares, or nanoflares) have been observed in multiple wavelengths including EUV and soft X-rays \citep{Kru1997, Par2000, Asc2000}. These brightenings have characteristic measured temperatures of 1--2~MK and derived energies of $10^{24}$--$10^{27}$~ergs. They release less energy, are shorter in duration, and occur much more frequently than X-ray bright points observed in the quiet Sun \citep{Gol1974,Kar2011}. \citet{Kru2000} observed this type of event using EUV emission measured by the Extreme Ultraviolet Imaging Telescope \citetext{EIT, \citealt{Del1995}} and the Coronal Diagnostic Spectrometer \citetext{CDS, \citealt{Har1995}} in addition to radio data from the Very Large Array (VLA). They concluded that quiet Sun heating events can be viewed as small flares, with similar temporal and spectral characteristics as larger events observed in active regions. The thermal components of such events may be difficult to detect with HXR instruments if their temperatures are low. On the other hand, if higher-temperature plasmas or significant non-thermal fluxes are present, these quiet Sun events could potentially be visible to HXR instruments more sensitive than {\rhessi}.  

The \textit{Nuclear Spectroscopic Telescope Array} ({\nustar}) uses focusing optics to directly image HXRs from $\sim$2 to 79 keV \citep{Har2013}. Though {\nustar} was designed as an astrophysics observatory, it can point at the Sun without any harm to the telescope optics and only a slight degradation in angular resolution \citetext{\citealt{Gre2016}, hereafter G16}. Here we perform the first search for transient, resolvable brightenings in quiet Sun regions observed by {\nustar}. We emphasize that these events are not the ``nanoflares'' referred to by modern theories of coronal heating \citep{Kli2015}, although {\nustar} can constrain the hot plasma they are predicted to produce \citep{Han2016}. We discuss the {\nustar} instrument and solar observing procedures in $\S$\ref{sec:paper1:observing}. Our analysis methods and results are described in $\S$\ref{sec:paper1:results}, and additional discussion of these results is found in $\S$\ref{sec:paper1:discussion}.

\section{Solar Observing with {\nustar}} \label{sec:paper1:observing}
{\nustar} is a NASA Astrophysics Small Explorer (SMEX) satellite launched on June 13, 2012 \citep{Har2013}. It has two co-aligned X-ray optics focused onto two focal plane detectors (FPMA and FPMB) and observes the sky in the energy range $\sim$2 to 79~keV. The instrument field of view (FoV) is approximately 12$'$$\times$12$'$ and the half-power diameter is $\sim$65$''$ \citep{Mad2015}. {\nustar} is well calibrated over the 3--79~keV bandpass and the lower energy bound can be extended to as low as 2.5 keV for spectroscopy if there is sufficient flux present (G16). \par

{\nustar} has been used to perform imaging spectroscopy on active regions \citep{Han2016}, to observe high-temperature loops after an occulted solar flare \citep{Kuh2017}, and to characterize sub A-class flares \citep{Wri2017, Gle2017}. The combination of a large effective area and low background rate makes it orders of magnitude more sensitive than {\rhessi}. This increase in sensitivity allows it to probe previously inaccessible regimes in flare parameter space, both in active regions and in the quiet Sun. However, because it was not designed to look at the Sun there are several limitations to {\nustar} that must be considered during solar observations planning and data analysis. \par
1) {\nustar} has a relatively low throughput of 400 counts s$^{-1}$ telescope$^{-1}$, which is reasonable for cosmic sources but very small for the Sun. Fortunately, this throughput limit is related only to digital data handling, and we can obtain data with minimal pileup at incident count rates as high as $\sim$$10^{5}$ counts s$^{-1}$ (G16). \par
2) Single-bounce photons from outside the FoV, known as ghost rays, can contribute significant emission inside the FoV \citep{Mad2015}. We have seen ghost ray patterns in several observations to date, and there is no easy way to remove this background. \par
3) The {\nustar} line-of-sight star tracker, or camera head unit (CHU), does not work during solar observing. There are three backup star trackers, all of which are oriented perpendicular to the instrument line of sight. As a result, offsets between the {\nustar} nominal and actual pointing can be as large as 1--2 arcminutes (G16). We must rely on direct comparisons with solar-dedicated imaging instruments such as \textit{SDO}/AIA to accurately calibrate our pointing. This is only possible when bright sources (e.g. active regions) appear in the {\nustar} FoV, and offsets are generally different for different CHU combinations. 

A full discussion of instrumental limitations and a summary of {\nustar} solar observations through April 2015 can be found in G16. Summary plots of all {\nustar} observations to date can be found at the dedicated website \url{https://ianan.github.io/nsigh_all/}.

\begin{figure*}[ht]
  \centering
   \begin{tabular}{ccc}
   \includegraphics[width=0.75\columnwidth]{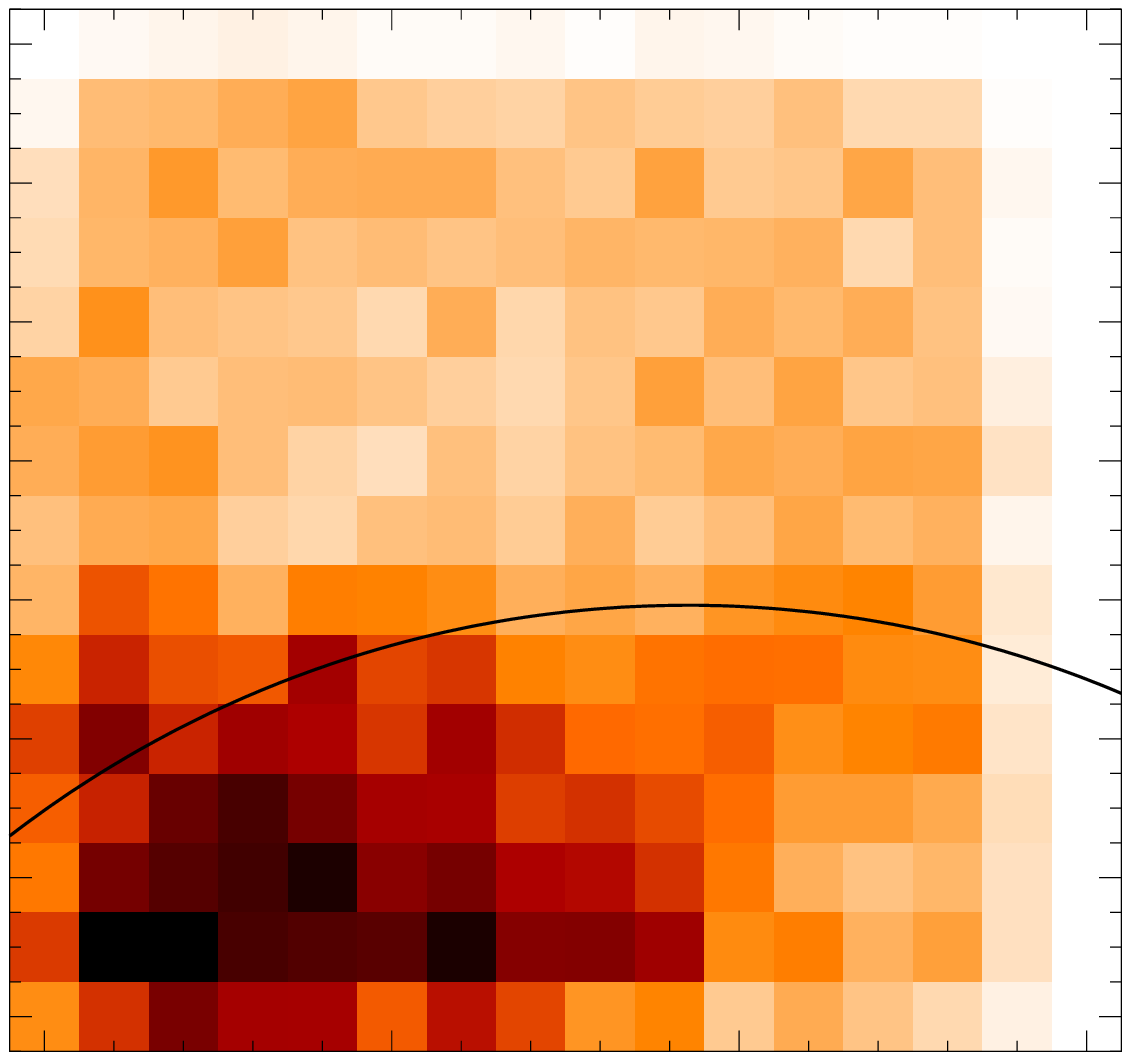}
   \includegraphics[width=\columnwidth]{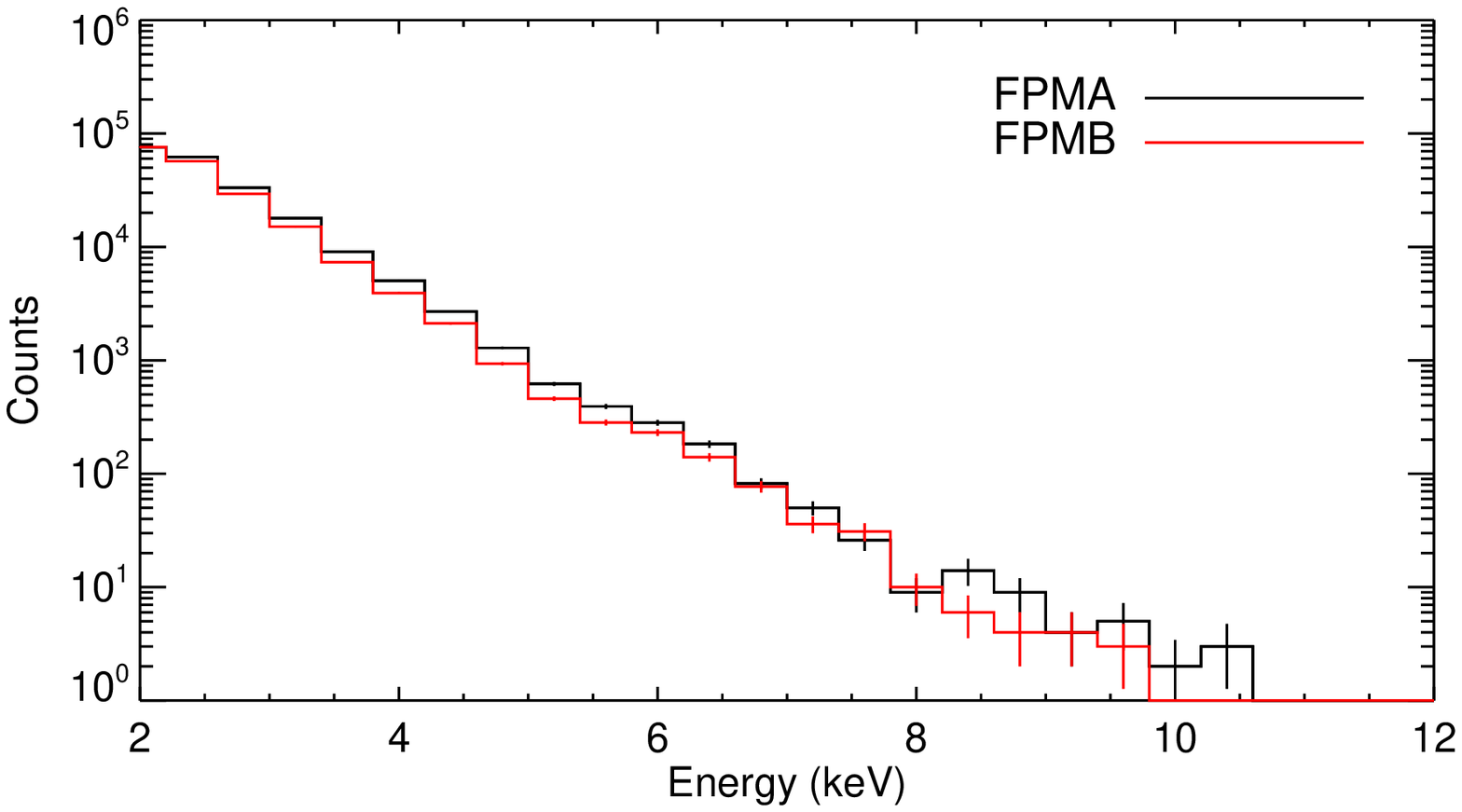}
   \end{tabular}
  \caption{(Left) Single frame of a {\nustar} FPMA image cube, with the solar limb overlaid in black. Spatial binning is 60$''$$\times$60$''$ and temporal binning is 100 seconds. (Right) {\nustar} count spectra from both telescopes, integrated over the full north pole pointing and the full FoV. Error bars shown are the square root of the number of counts in each bin. Most or all of the counts in both panels are due to ghost rays from active regions outside the FoV.}
  \label{fig:imcube}
\end{figure*}

\section{Analysis and Results} \label{sec:paper1:results}
\subsection{Data reduction}
The data presented in this paper are from the fourth orbit of the second {\nustar} solar campaign, which took place on 2014 November 1. This orbit included two quiet Sun pointings and the lowest solar flux levels of this campaign (full Sun \textit{GOES} class $\sim$B4). We analyzed data from the second quiet Sun pointing (aimed at the north pole) due to a reduced ghost ray background. The active regions observed during the first two pointings in this orbit are studied in \citet{Han2016}.

Event files were generated and processed using the {\nustar} Data Analysis software v1.4.1 and {\nustar} calibration database 20150414. We utilized a modified analysis pipeline for solar data, as the standard pipeline throws out a large fraction of real events (G16). The {\nustar} event files were translated into heliocentric coordinates using the JPL Horizons\footnote[1]{http://ssd.jpl.nasa.gov/horizons.cgi} database of solar RA/DEC positions. Non-physical events (e.g. photons with zero energy, uncalibrated positions, or in hot pixels) were thrown out. In addition, we estimated the fraction of piled-up events by taking the ratio of events with non-physical grades to the total number of events (G16). A ``grade'' is a number assigned to every event based on which pixels in a 3$\times$3 grid centered on a triggered pixel collect charge above a certain threshold. We found negligible pileup fractions of 0.067\% and 0.062\% for FPMA and FPMB, respectively. To further ensure that our results were unaffected by pileup, we threw out events with non-physical grades. 

We used \textit{SDO}/AIA data to calibrate the {\nustar} pointing alignment. All of the data for the north pole pointing were taken in CHU combination 1+3. Fortunately, the same CHU combination was used in a previous orbit during active region observations. We used active region pointings in consecutive orbits to verify that the offsets of different CHU states stayed approximately the same from orbit to orbit. A shift of (x-105$''$, y+65$''$) applied to the {\nustar} images gave the best match to active region positions measured in CHU state 1+3.  

Figure \ref{fig:profiles} shows {\nustar} counts $>$2~keV in the FPMA telescope integrated over the full north pole pointing. There are more counts on disk than off disk, but we were unable to unambiguously distinguish the solar limb. We therefore cannot claim a definitive detection of HXR emission from the quiet Sun on this basis. There are $\sim$14 minutes of data between the times {\nustar} entered sunlight and entered the South Atlantic Anomaly (seen as a livetime dropout at $\sim$22:28 UT in Fig. \ref{fig:profiles}). Though the Sun was mostly quiet during this pointing, a small microflare occurred near disk center at $\sim$22:18 UT and is visible in the {\rhessi} light-curve. Ghost rays from this event are correlated with a decrease in livetime, although the effect is no larger than variations during non-flaring periods. However, the ghost ray contribution from this microflare resulted in significant brightenings that will be discussed in Section \ref{search}. 

We generated 3-D image cubes by binning the {\nustar} event files in space and time.  Figure \ref{fig:imcube} shows a single frame of the FPMA image cube, with the solar limb overlaid in black and binning of 100s and 60$''$$\times$60$''$. This image includes the pointing correction derived from \textit{SDO}/AIA data. This figure also shows the integrated count spectrum in both telescopes for the north pole pointing. {\nustar} does not see any counts $>$11~keV, though we can set flux limits at lower (2.5 to 4~keV) and higher (10 to 20~keV) energies. Our particular choices of spatial and temporal bins are discussed in Sections \ref{lowe} and \ref{highe}. 

Simulations with the full-instrument simulator NuSIM \citep{Mad2011} showed that the observed quiet Sun emission is consistent with ghost rays produced by active regions near Sun center, outside the instrument FoV. We can compare the observed flux with the {\nustar} non-solar background to determine its importance. The {\nustar} background spectrum at energies $<$20~keV is dominated by the ``aperture'' component, or stray light from outside the instrument FoV that can shine directly on the detectors \citep{Wik2014}. Close to 20~keV there is a nearly equal contribution from internal scattering. When {\nustar} is pointed at the Sun the magnitude of the ``solar'' background component (a result of sunlight backscattering off the mast and optics) goes to zero. We estimated the approximate number of background counts in each telescope from 2.5 to 4~keV and 10 to 20~keV using the \citet{Wik2014} blank sky fluxes at 3 and 10~keV respectively. Over the duration of the quiet Sun observations (27.7 s exposure, from 800 s observing time at 3.5\% livetime) the blank sky data predicts 0.166 counts telescope$^{-1}$ from 2.5 to 4 keV and 0.44 counts telescope$^{-1}$ from 10 to 20 keV. In comparison the integrated number of observed counts in each telescope was approximately 10$^{5}$ for the low energy band and 5 for the high energy band. Therefore the instrumental background is unimportant at low energies and unlikely to lead to significant statistical fluctuations at higher energies. 

\begin{figure}[t]
    \centering
    \includegraphics[width=0.31\columnwidth]{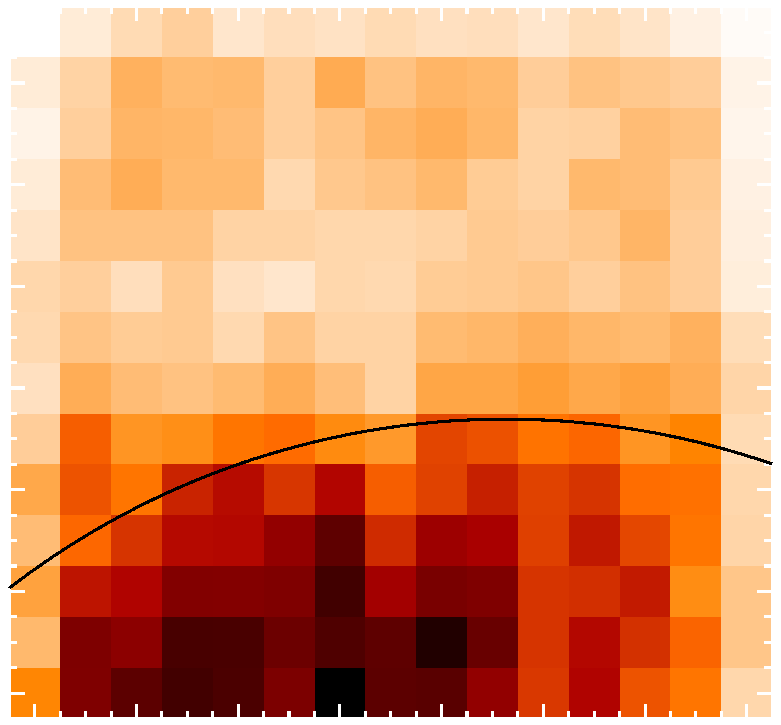} 
    \includegraphics[width=0.31\columnwidth]{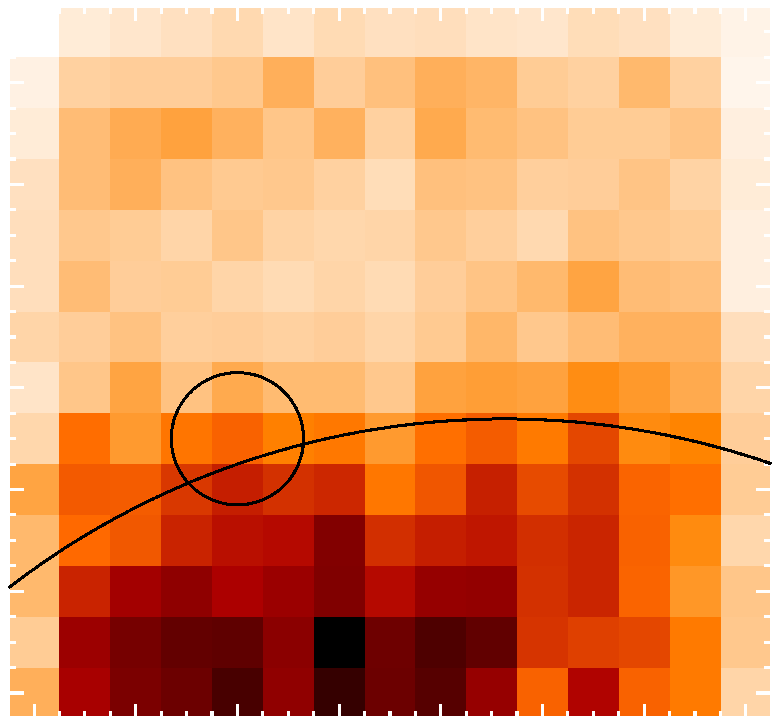}
    \includegraphics[width=0.31\columnwidth]{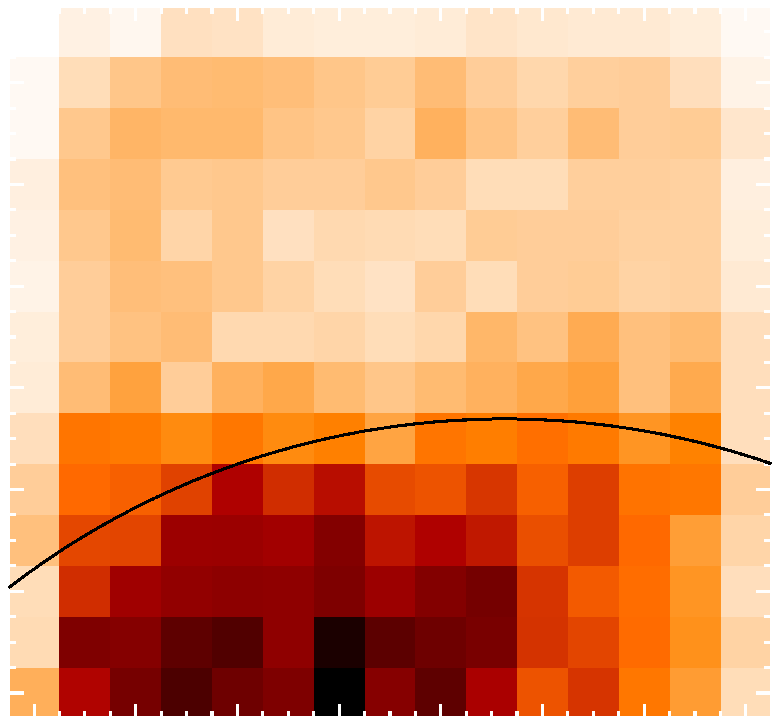}  
    \includegraphics[width=0.31\columnwidth]{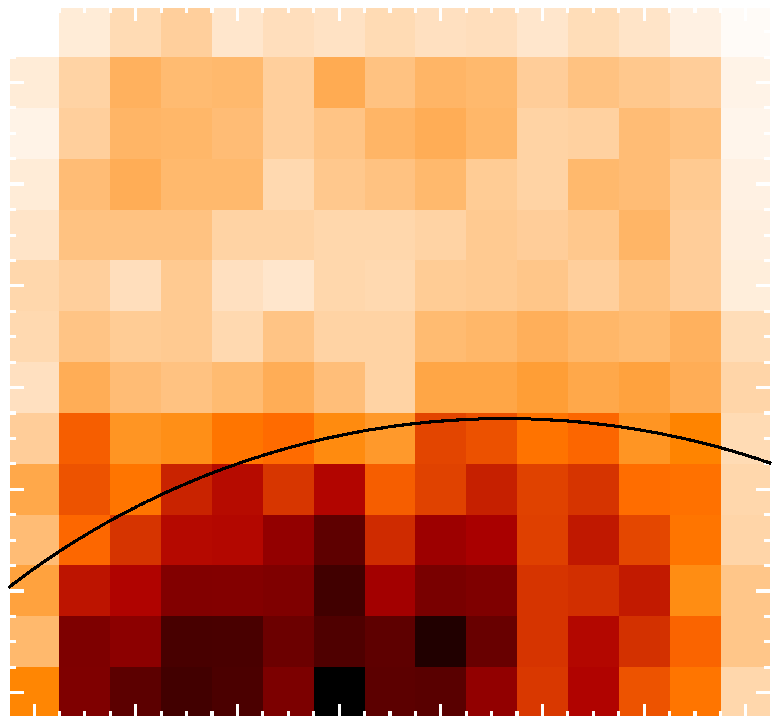} 
    \includegraphics[width=0.31\columnwidth]{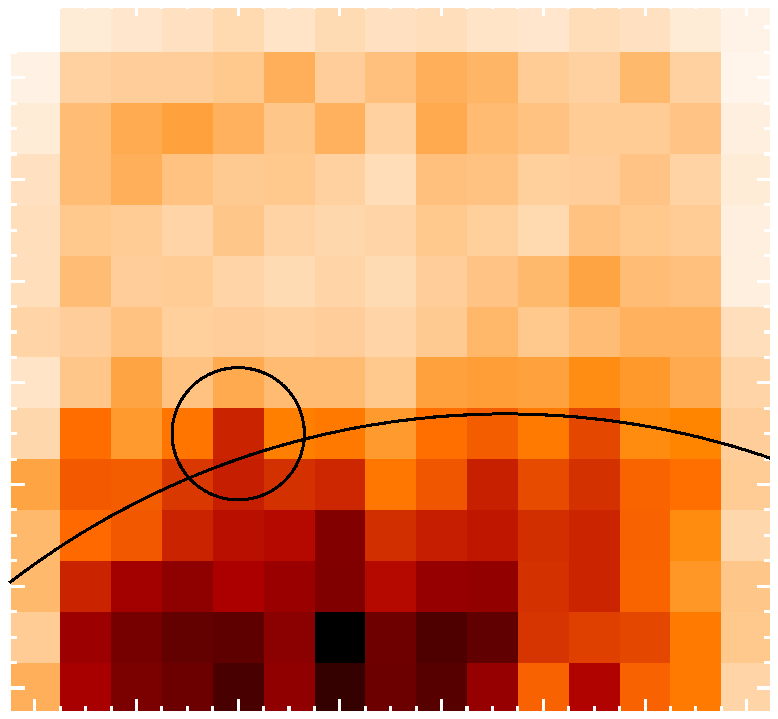} 
    \includegraphics[width=0.31\columnwidth]{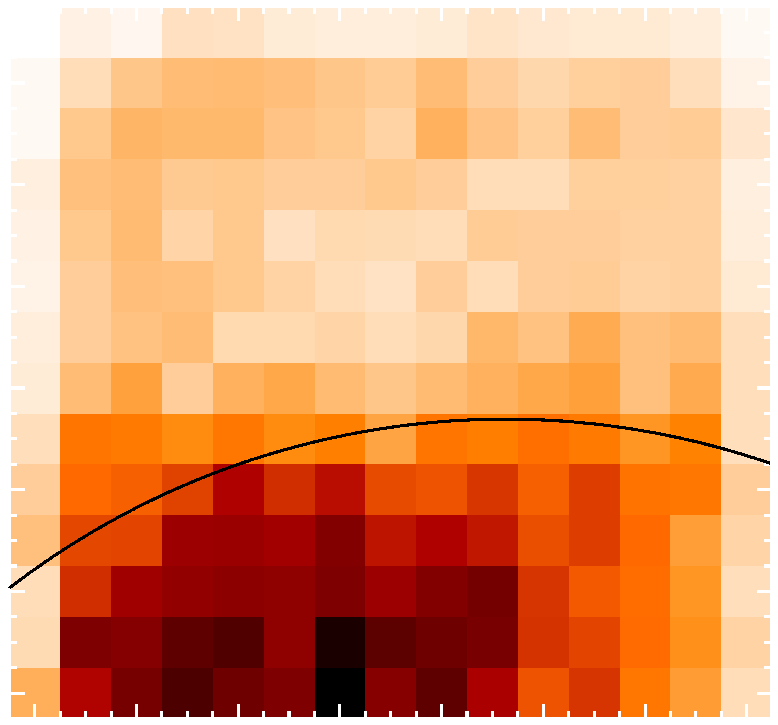} 
\caption{Temporally adjacent {\nustar} image cubes with 100 s temporal and 60$''$$\times$60$''$ spatial binning, with time increasing from left to right. The solar limb is marked by a black line. (Top) Original image cubes with a ``source'' pixel marked by a black circle. (Bottom) The central image cube has just enough excess counts added to the source pixel to reach the 95\% detection threshold.}\label{fig:example}
\end{figure}

\begin{figure*}[th]
  \centering
   \includegraphics[width=\columnwidth]{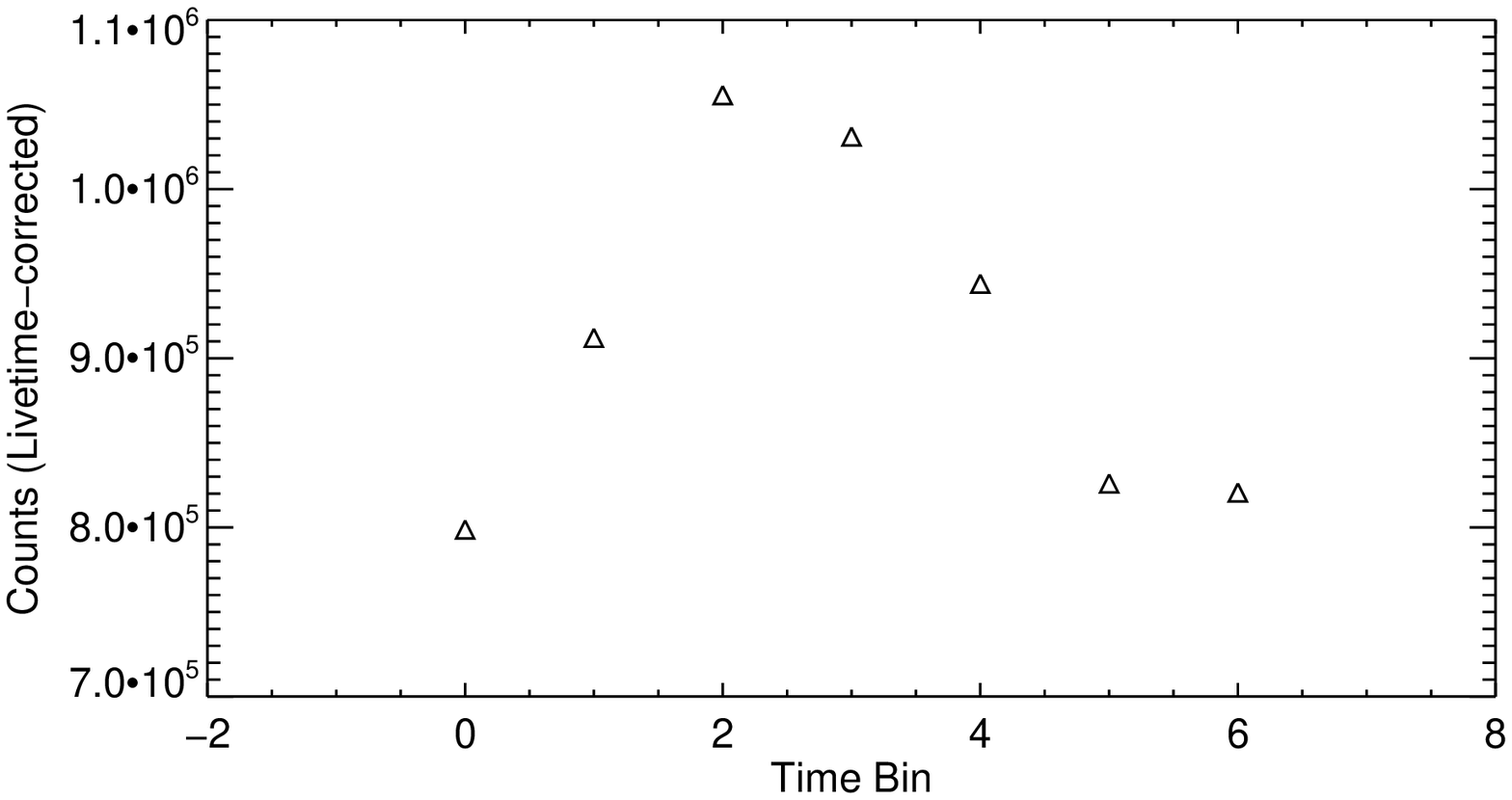}
   \scalebox{0.7}{\includegraphics[width=\columnwidth]{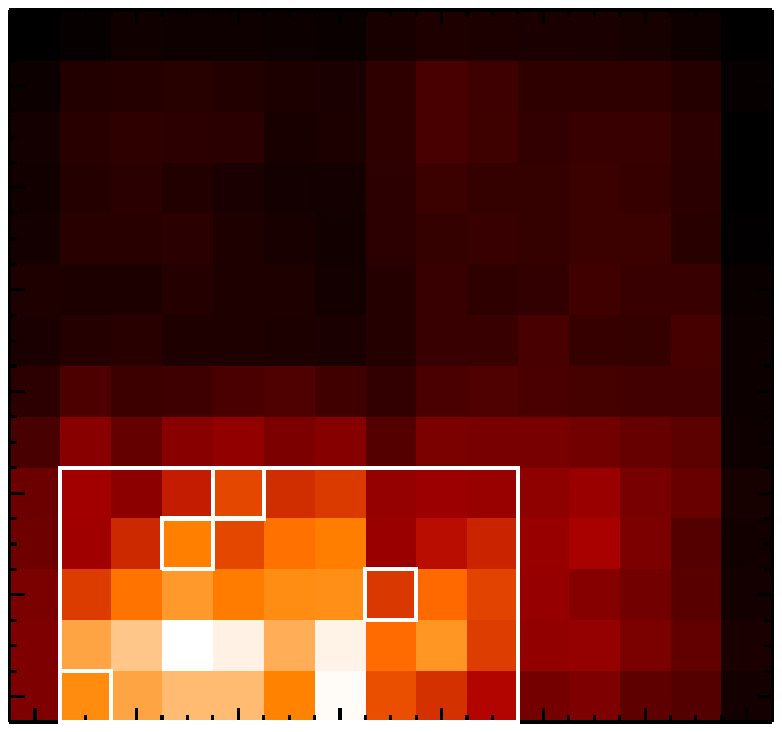}}
  \caption{(Left) Incident counts during this observation per 100 s time bin for the region shown in the right panel. (Right) Bright region of the {\nustar} FoV selected for the light-curve plotted in the left panel indicated by a white rectangle. The pixels outlined in white were not included in the light-curve due to the presence of fluxes above the detection threshold in the peak time bin.}  
\label{fig:transients}
\end{figure*}

\subsection{Adding the telescopes} \label{adding}
{\nustar} has two focal planes (FPMA \& FPMB) with a throughput limit of 400 counts s$^{-1}$ each. Although the telescopes are read out separately, the data for both can be combined with care. This is desirable because if there is a real signal anywhere in our time series, doubling the signal and background by adding the telescopes will gain us a factor of $\sqrt{2}$ in the signal-to-noise ratio. However, we can only add the telescopes if their spatial differences are negligible. Because of the spacecraft geometry, the ghost ray patterns can be different in each telescope for the same FoV. In extreme cases the sensitivity in a particular region can be much better in one telescope than in the other, as the result of a reduced ghost ray background.

For a given spatial pixel at a given time, if the ratio of the higher number of counts to the lower number of counts in each telescope is $>$3 the gain in signal-to-noise ratio is negated by adding a quiet pixel to a noisy one. However, if there are fewer than 10 combined counts in a given spatial pixel, there is a large uncertainty in the counts ratio. Therefore, we used the sum of both telescopes for every pixel \emph{unless} both of these conditions were met: the number of summed counts $>$10 and the ratio of the higher number of counts to the lower number of counts is $>$3. Combined pixels comprised about 94\% of those used for analysis; the remaining pixels were taken from one of the two telescopes. For summed pixels we used the average of the two telescope livetimes. 

\subsection{Probability Calculations} \label{probability}
We used Poisson statistics to determine the probability of getting \textit{S} or more counts in a particular macropixel given a background \textit{B}. This tests the null hypothesis that $S$ is from background alone in the absence of any signal. We calculated the background by averaging counts in the same spatial macropixel in adjacent temporal frames, accounting for changes in livetime. If a frame was the first or last of the image cube, then we used the single temporally adjacent macropixel as the background. The one and two frame background equations, respectively, are as follows: 

\begin{equation}
  B = L_{t} \frac{N_{t\pm1}}{L_{t\pm1}}
\end{equation}
\begin{equation}
  B = \frac{1}{2}L_{t} \left( \frac{N_{t-1}}{L_{t-1}} + \frac{N_{t+1}}{L_{t+1}} \right)
\end{equation}
where $N_{t}$ and $L_{t}$ are the number of counts and the livetime, respectively, in the $t^{th}$ frame. Since we do not know $\lambda$ (the true background rate of which $B$ is a sample) a priori, we generated databases of cumulative Poisson probabilities for a wide range of ``source'' and ``background'' counts and for 1 and 2 background frames. Given $S$ source counts in the pixel of interest and an average background $B$, we computed the cumulative Poisson probability $P_{\ge}(S)|_{B}$ as follows. 

First we generated a large number of trials for source ($S$) and background ($B$) counts using a Poisson distribution with average value $\lambda$. For the low energy (2.5 to 4 keV) image cube we used a range of 0.5 to 1200 counts macropixel$^{-1}$ for $\lambda$, with a spacing of 0.5 counts macropixel$^{-1}$. This range was chosen to include values of $\lambda$ up to 2.5 times the maximum value in a single macropixel. We then created a 3-D array with each element equal to the number of occurrences of [$B$,$S$,$\lambda$], and summed over the third dimension of this array to marginalize $\lambda$. The rows of the resulting 2-D array were normalized so that each had unit sum. This set the probability of getting any value of $S$ for a particular value of $B$ to 1 (as it should be). The last step was to integrate all probabilities $\ge$$S$ for each location [$B$,$S$] in the databases, yielding the probability of getting $S$ or more counts for a given $B$.

After we generated the Poisson databases, we calculated the cumulative probability $P_{\ge}(S)|_{B}$ for every pixel in every time frame of the binned, combined image cube. We analyzed only pixels with their center on the solar disk, and performed the calculations on two image cubes with different time bins: the default bins and the default bins shifted half a bin forward in time. The purpose of the temporal shift was to increase the sensitivity to events that occurred on or near the default bin edges. Time bins with only partial data coverage were thrown out.

\subsection{Transient Search} \label{search}
We chose initial spatial and temporal binnings for the transient search of 60$''$$\times$60$''$ and 100 seconds. The spatial binning is approximately the instrument half-power diameter (within which half the flux of a point source is expected to fall). The temporal binning is a duration that should be longer than an appreciable fraction of faint, transient HXR events. The average duration of a sample of microflares seen by {\rhessi} is $\sim$6 minutes, and the shortest events in that sample are $\sim$1 minute long \citep{Chr2008}. We expect {\nustar} to be sensitive to events at least an order of magnitude fainter than those seen by {\rhessi}, with correspondingly shorter durations; see e.g. \citet{Ver2002} for the correlation of flare duration with X-ray flux. 

\begin{figure*}[ht]
  \centering
   \begin{tabular}{ccc}
   \includegraphics[width=\columnwidth]{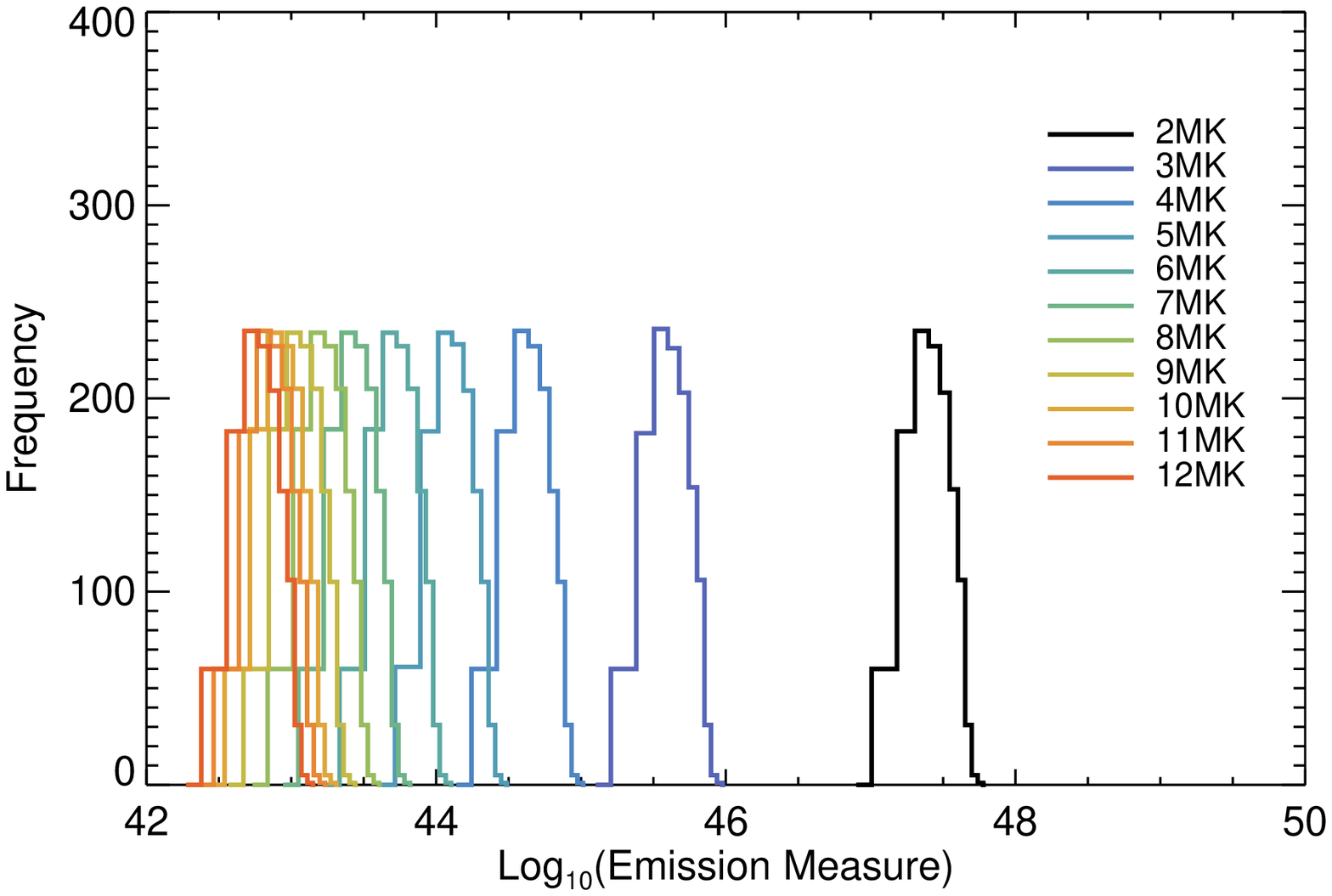}
   \includegraphics[width=\columnwidth]{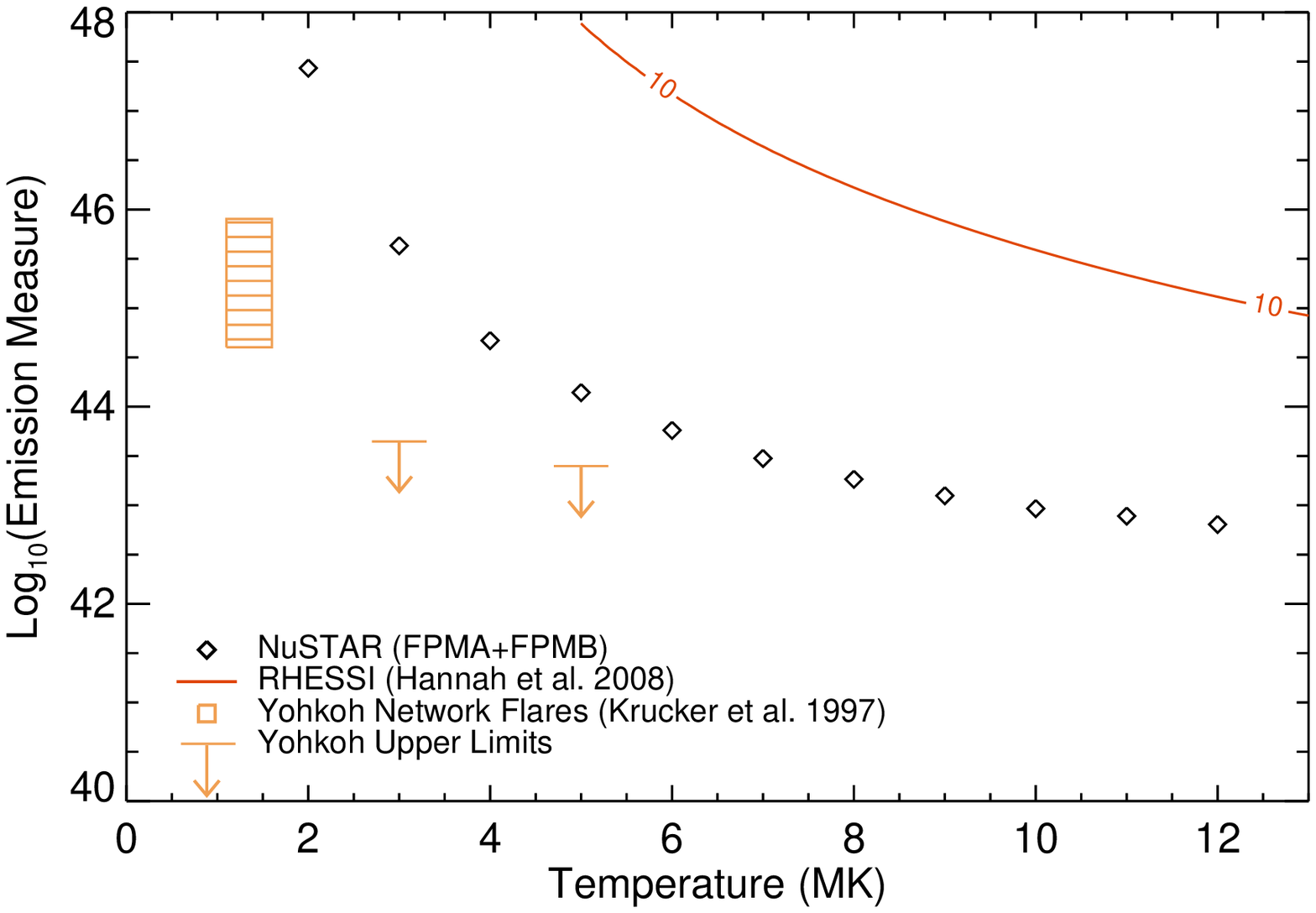}
  \end{tabular}
  \caption{(Left) The ``just detectable'' emission measure distributions for temperatures 2--12~MK, calculated for summed (FPMA+FPMB) north pole image cubes. These distributions include every macropixel from two image cubes: one with no time shift and one with a half-bin time shift. The ``just detectable'' limit corresponds to the intensity that gives a count excess above background at the 95\% confidence level. (Right) The {\nustar} sensitivity for this observation with $t_{bin} = 100$s and $s_{bin}=$60$''$$\times$60$''$. The black diamonds correspond to the peaks of the EM distributions in the left plot. The red curve is the level at which {\rhessi} would detect 10 cts s$^{-1}$ detector$^{-1}$, approximately the instrument limit for imaging and spectroscopy. The quiet Sun transient brightenings observed by \textit{Yohkoh}/SXT \citep{Kru1997} are shown as an orange striped box; these events are below the sensitivity limit for this observation. \textit{Yohkoh}/SXT upper limits on higher-temperature brightenings are shown as orange arrows.}  
\label{fig:lowe}
\end{figure*}

In order to search for transient events we calculated the probability of observing $S$ counts in a particular pixel given the average background $B$ in the temporally adjacent pixel(s). For upper limit calculations (Sections \ref{lowe} and \ref{highe}) we determined the minimum number of counts $S$ necessary to meet the 95\% confidence detection threshold, accounting for the number of trials. These two cases are shown in Figure \ref{fig:example} for one particular 60$''$$\times$60$''$ pixel. These plots show binned image cubes for three consecutive 100 s dwells (time increases from left to right), with the example pixel is indicated by a black circle. Two cases are shown: the original, livetime-corrected image cubes (top) and the same image cubes where the number of counts in the example pixel has been increased to meet the detection threshold (bottom). With a 100 s dwell and 60$''$$\times$60$''$ macropixels there are 210 spatial pixels and 7 time bins, for a total of 1470 spatiotemporal pixels. Therefore the probability threshold is 0.05$/$1470$/$2 = 1.7$\times$$10^{-5}$; smaller event probabilities are required for a detection at the 95\% confidence level. The extra factor of 2 is a conservative way to account for the half-bin temporal shift (conservative because the shifted pixels are not totally independent of the un-shifted pixels, so the exact correction factor for the number of trials would be $<$2). 

The average livetime-corrected background $B$ for the source pixel in Figure \ref{fig:example} is 132 counts, and the original number of source counts $S$ is 144. In order to meet the 95\% threshold 198 counts are required. Note that the Poisson probability of seeing $\ge$198 counts, given an expected value of 132, is 5.2$\times$10$^{-8}$. However, this fails to take into account the uncertainty in the true background value $\lambda$. Our more comprehensive calculation, which includes integrating over a wide range of $\lambda$, gives a probability of 1.3$\times$10$^{-5}$ (just below the threshold for this image cube). In other words, a proper handling of the background uncertainty increases the probability of this event (and of higher count rates in general), making it more difficult to claim a detection. Note that the sensitivity can vary strongly depending on the background level in a particular pixel, since a significant transient in one pixel is much fainter than the background levels in others.

We conducted a transient search over the full {\nustar} energy range and over a low energy band of 2.5--4~keV. There are several pixels with brightenings above the 95\% confidence limit, all in the same time bin (22:14:42 - 22:17:22 UT) in non-adjacent pixels. The full light-curve for the bright region of the chip (excluding the pixels above threshold) is shown in Figure \ref{fig:transients}. The bin with the highest (livetime-corrected) count rate (Figure \ref{fig:transients}) is the bin in which all the transients were detected. This corresponds to the rise phase of a microflare near disk center; this event caused a visible increase of the ghost-ray flux during these observations. Furthermore, an examination of co-spatial and co-temporal \textit{SDO}/AIA images did not find any transient events. The combined evidence indicates that these detections are a result of the microflare outside the FoV and are not transient events in the quiet Sun.

In conclusion, we have insufficient evidence of any transient brightenings in the quiet Sun on time scales of 100 s. We can use quiet Sun transient observations by e.g. \textit{Yohkoh}/SXT \citep{Kru1997} to determine the likelihood of seeing similar phenomena during this {\nustar} pointing. \citet{Kru1997} calculated an occurrence rate of 1 event every 3 s over the entire solar surface, which corresponds to 22 events over the duration and FoV of the 2014-November-01 quiet Sun observation. However, as we shall see in the following sections, the {\nustar} sensitivity is insufficient to detect events similar to those observed by \textit{Yohkoh}/SXT.

\begin{figure*}[ht]
  \centering
  \scalebox{0.6}{\includegraphics{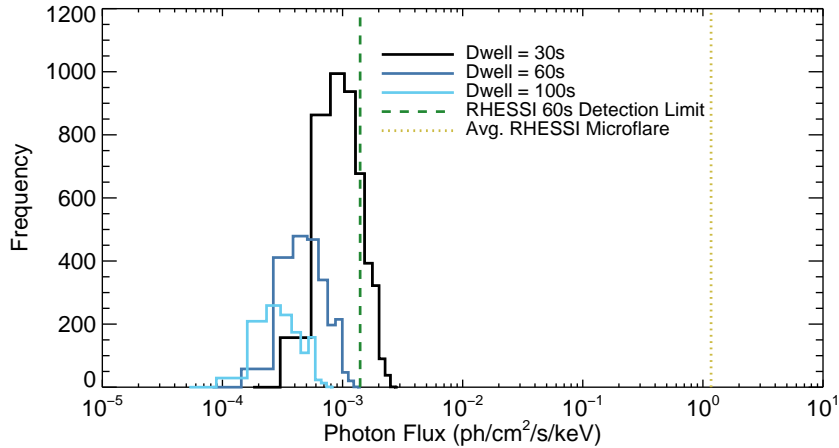}}
  \caption{{\nustar} limits on 10--20~keV photon flux for this observation, calculated for the sum of FPMA \& FPMB and three different temporal binnings (dwells). Each distribution includes every macropixel from two image cubes: one with no time shift and one with a half-bin time shift. The dashed line is the {\rhessi} detection limit at 10~keV. The dotted line is the average {\rhessi} microflare flux at 10~keV from \citet{Han2008}.}
  \label{fig:highe}
\end{figure*}

\subsection{Low Energy (Thermal) Limits} \label{lowe}
In the absence of any definitive transient detections, we chose an energy range 2.5--4~keV to place upper limits on thermal emission. In this range the {\nustar} instrument response is well understood and there were a relatively large number of counts. We chose the same temporal and spatial binnings used for the transient search (100 s and 60$''$$\times$60$''$). 

We generated isothermal bremsstrahlung spectra with temperatures 2--12~MK using the {\tt f\_vth.pro} function in Solarsoft \citep{Fre1998}. Next we converted the spectra from photons to counts using the {\nustar} diagonal response matrix (used hereafter), assuming a point source and correcting for livetime. Non-diagonal effects such as K-shell escape peaks should be negligible due to the steeply falling nature of the observed spectrum. The only relevant peak in the CdZnTe detectors is Zn (characteristic energy $\sim$8.5~keV) and the observed {\nustar} counts flux decreases by roughly four orders of magnitude as energy increases from the lower limit of instrumental sensitivity to 8.5 keV higher. After determining the peak number of counts produced in a macropixel by each isothermal spectrum, we calculated the number required to meet the probability threshold based on the background level of each macropixel. We again performed the half-bin forward time shift, and accumulated statistics over two separate image cubes. Each image cube was a combination of FPMA and FPMB determined by the criteria in $\S$\ref{adding}.   

The {\nustar} detectors are subject to vignetting as a function of off-axis angle \citep{Mad2015}. We used instrument vignetting curves from the {\nustar} calibration database to adjust the count thresholds. For each 60$''$$\times$60$''$ macropixel, the average off-axis angle of every event was calculated and the vignetting curve function for the closest tabulated angle (averaged over the energy range 2.5--4~keV) was used as a correction factor. In addition, we calculated the fractional flux from a point source contained in a 60$''$$\times$60$''$ macropixel and multiplied the number of counts by this factor.

After we applied the vignetting corrections to the count detection thresholds in each macropixel, we divided the counts by the temperature response functions from 2 to 12~MK. In this way we obtained the emission measures of isothermal spectra for every macropixel. The left panel of Figure \ref{fig:lowe} shows the distribution of emission measures at the detection threshold for isothermal temperatures between 2 and 12~MK. The right panel of Figure \ref{fig:lowe} shows the {\nustar} T and EM sensitivity curve from 2 to 12~MK. Each black diamond corresponds to the peak of the EM upper limits distribution for a particular temperature. Plotted in red is the {\rhessi} 10 counts s$^{-1}$ detector$^{-1}$ contour; this is approximately the lowest count rate at which {\rhessi} can perform imaging and spectroscopy. {\nustar} is sensitive to events 2--4 orders of magnitude smaller than the smallest microflares seen by {\rhessi}. The T and EM range of quiet Sun network flares seen by \textit{Yohkoh}/SXT is shown by an orange box \citep{Kru1997}. We also calculated \textit{Yohkoh}/SXT upper limits on hotter network flares with temperatures of 3 and 5 MK (orange arrows). While {\nustar} is not sensitive enough at low temperatures to detect quiet Sun brightenings similar to those observed by \textit{Yohkoh}/SXT, future observations with higher livetime could allow {\nustar} to detect high-temperature or non-thermal components of those events if they exist.  

\subsection{High Energy Limits} \label{highe}
Important physical mechanisms such as impulsive heating and particle acceleration can be constrained by the presence (or absence) of nonthermal emission at energies $>$10 keV. Therefore, we calculated 10--20~keV photon flux limits based on the {\nustar} quiet Sun observations. We used a procedure similar to the one used to calculate the low energy limits, modified to account for the low statistics present in this energy range. {\nustar} saw a total of 15 counts between 10 and 20~keV in 801 seconds of north pole observing time, but no more than one count in any $60''\times60''$, 100 s macropixel (even after adding both telescopes using the criteria in $\S$\ref{adding}). Because the number of counts is so small, we could strongly constrain the true background count rate $\lambda$. We set a conservative upper limit on $\lambda$ of two times the average number of counts per macropixel in the on-disk portion of the image cube. We consider this conservative because in this energy range the background is dominated not by ghost rays, which have a lot of spatial structure, but by the relatively uniform instrumental background.  The results are, in fact, very sensitive to the range of $\lambda$ since the values of $\lambda$ near the cutoff dominate the marginalized probability when $B=0$ (as it was for most macropixels). We calculated cumulative probabilities, summed over this limited range of $\lambda$, to determine the number of counts required to reach the 95\% threshold. We then converted from counts to photons using the {\nustar} effective area, livetime correction, vignetting coefficients, and the fractional flux contained in a macropixel. We used the effective area at 10 keV for two reasons. First, while these limits are model-independent most microflare nonthermal spectra are steeply falling. In addition, {\nustar}'s effective area varies little between 10 keV and 20 keV. 

\begin{figure*}[ht]
  \centering
  \scalebox{0.6}{\includegraphics{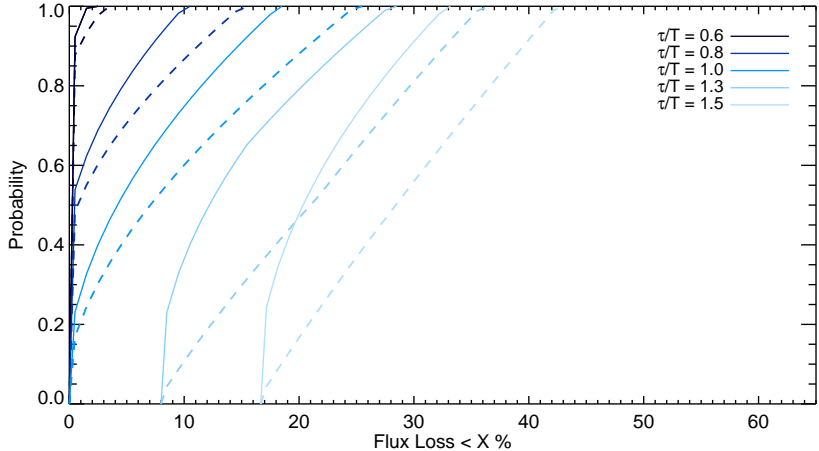}}
  \caption{Cumulative probability distributions of flux loss for several different values of flare duration divided by bin width. For values of $\tau/T < 0.5$ there is no flux loss in the best time bin, which we selected for every trial. Results are shown for a triangle profile (solid lines) and a half-triangle profile with an immediate rise and linear decay (dashed lines).}
  \label{fig:fluxloss}
\end{figure*}

We explored shorter time bins for this energy range because higher-energy transient X-ray emission (such as seen during flares) is generally shorter-lived. Figure \ref{fig:highe} shows the {\nustar} flux limits from 10 to 20~keV for temporal binnings (dwells) of 30, 60, and 100 s. On the same graph are the average {\rhessi} microflare flux \citep{Han2008} at 10~keV and the {\rhessi} 60 second detection limit in this energy range. For the latter we assumed a 10--20~keV background of 1 count s$^{-1}$ detector$^{-1}$ and 9 live detectors, then calculated the incident flux necessary for a 95\% detection. In this energy range {\nustar} is sensitive to transient events several times smaller than the {\rhessi} detection threshold for identical integration times. While this is a useful comparison, note that lower-energy bands are better for source detection with both instruments due to higher flux. 

\section{Discussion} \label{sec:paper1:discussion}
\subsection{Event Duration and Flux Loss}
For the transient search ($\S$\ref{search}), we searched consecutive time intervals and also the same set of intervals shifted ahead by half the interval duration, to ensure that we captured as much signal as possible for a random brightening. However, for event durations $\tau$ that are an appreciable fraction of a time bin width $T$, there will be flux loss even in the bin that contains most of the event. Flux loss means that some fraction of counts from a transient event falls outside the time bin of interest. This results in signal lost and additionally subtracted from the remaining flux, since it spills over into a time interval used for background subtraction. The amount of flux lost depends on the event duration, which can be conveniently expressed as a fraction of the bin length ($\tau/T$), and on the (arbitrary) start time of the transient relative to the time bin edges. 

To quantify the average flux loss for a particular value of $\tau/T$, we performed Monte Carlo simulations with a series of time bins and randomly injected transient events. For each event we examined both unshifted and shifted time bins and selected the bin with the least amount of flux lost. Then we generated a flux loss probability distribution based on a large number of simulations. Figure \ref{fig:fluxloss} shows the cumulative distribution of the amount of flux loss for two flare shapes (triangle and half-triangle) for a range of flare durations (expressed as a fraction of the bin duration). 

A triangular profile results in lower flux losses than a half-triangle profile of the same duration, as expected. For a triangular profile, the maximum flux loss is about 33 percent for the longest duration flares ($\tau/T$=1.5). We did not analyze values of $\tau/T$$>$1.5, as events that long would have to be extremely bright to be detected by the search algorithm. For events shorter than a time bin, the triangular events exhibit $<$15\% flux losses. We do not expect this amount of loss to have a significant effect on our results; detailed simulations of the effects on transient sensitivity will be discussed in a future paper. 

\subsection{Future Observations}
\textit{Yohkoh}/SXT was sensitive to a nominal energy range of $\sim$0.25--4~keV, about an order of magnitude lower in photon energy than {\nustar}. At these energies solar fluxes are much higher and lower temperatures dominate the emission. Therefore {\nustar} is not sensitive enough to detect brightenings similar to the network flares seen by \citet{Kru1997}. However, the gain in sensitivity at higher energies was significant compared to previous HXR observations (as seen in Figure \ref{fig:lowe}). In addition, we expect higher sensitivity in future observations due to decreasing solar activity. This increase will result from two factors: higher livetime due to lower incident count rates, and a decrease in the solar background count rate from the maximum throughput level to as low as the level of the quiet corona. For this observation the average incident 2.5--4~keV background (ghost-ray) count rate was $\sim$16 counts s$^{-1}$ arcmin$^{-2}$ telescope$^{-1}$; this is simply the average count rate per macropixel in this energy range, corrected for livetime. In comparison, the estimated incident rate from the quiet corona at solar minimum \citetext{spectrum from \citealt{Syl2010}} is $\sim$0.98 counts s$^{-1}$ arcmin$^{-2}$ telescope$^{-1}$ in this energy range. This rate was derived by multiplying the average 2.5--4~keV flux from \citet{Syl2010} (units of photons cm$^{-2}$ s$^{-1}$ keV$^{-1}$) by one {\nustar} telescope's effective area ($\sim$90~cm$^{2}$), the energy range width (1.5~keV), and a factor of 1/144 to scale from the full FoV to a square arcminute. The count rate discrepancy between this observation and the quiet Sun at solar minimum is even greater above 4~keV. A similar calculation gives an expected count rate from 4 to 20~keV of $\sim$3.9$\times$10$^{-5}$ counts s$^{-1}$ arcmin$^{-2}$ telescope$^{-1}$ for the \citet{Syl2010} spectrum, over $\sim$4 orders of magnitude smaller than the count rate seen in this observation (1.7 counts s$^{-1}$ arcmin$^{-2}$ telescope$^{-1}$). At that level of activity, non-solar background would be the dominant source of high-energy emission in the {\nustar} FoV; the blank-sky spectra from \citet{Wik2014} give incident background rates of $\sim$4$\times$10$^{-5}$ counts s$^{-1}$ arcmin$^{-2}$ telescope$^{-1}$ in a narrow band from 2.5 to 4 keV, and $\sim$2$\times$10$^{-4}$ counts s$^{-1}$ arcmin$^{-2}$ telescope$^{-1}$ in a wider band from 4 to 20 keV. The \citet{Syl2010} measurements were made during unusually low levels of solar X-ray activity, so we anticipate 2--3 orders of magnitude increased sensitivity with {\nustar} in the current cycle.

\null

\section*{Acknowledgements}
This paper made use of data from the {\nustar} mission, a project led by the California Institute of Technology, managed by the Jet Propulsion Laboratory, and funded by the National Aeronautics and Space Administration. We thank the {\nustar} Operations, Software and Calibration teams for support with the execution and analysis of these observations. This research has made use of the {\nustar} Data Analysis Software (NUSTARDAS) jointly developed by the ASI Science Data Center (ASDC, Italy) and the California Institute of Technology (USA). Some of the figures within this paper were produced using IDL color-blind-friendly color tables \citep{Wri2017b}. The authors would like to thank Albert Shih for helpful comments and suggestions. AM was supported by NASA Earth and Space Science Fellowship award NNX13AM41H. IGH acknowledges support from a Royal Society University Research Fellowship. SK acknowledges support from the Swiss National Science Foundation (200021-140308). AC was supported by NASA grants NNX15AK26G and NNX14AH54G. LG was supported by an NSF Faculty Development Grant to UMN (AGS-1429512). PJW was supported by an EPSRC/Royal Society Fellowship Engagement Award (EP/M00371X/1). This work was supported by NASA grants NNX12AJ36G and NNX14AG07G. 

\facility{NuSTAR}.

\bibliography{nustar_paper_draft10_aas}

\clearpage

\end{document}